\documentclass{baltzer}
\usepackage{graphics}
\begin{document}
\begin{frontmatter}  
%
%\pretitle{}                  %   e.g.: \pretitle{Guide}  
\title{Precise Measurement of Muon Capture on the Proton\thanks{work supported by
Paul Scherrer Institute, Russian Ministry of Science and Technology, 
US Department of Energy, US National Science Foundation and INTAS.}  }
\author[4]{P. Kammel}
\author[1]{V.A. Andreev}
\author[1]{D.V. Balin}
\author[7]{R.M. Carey}
\author[3]{T. Case}
\author[4]{D.B. Chitwood}
\author[4]{S.M. Clayton}
\author[3]{K.M. Crowe}
\author[5]{J. Deutsch}
\author[4]{P.T. Debevec}
\author[2]{P.U. Dick}
\author[2]{A. Dijksman}
\author[2]{J. Egger}
\author[2]{D. Fahrni}
\author[1]{A.A. Fetisov}
\author[3]{S.J. Freedman}
\author[1]{V.A. Ganzha}
\author[3]{B. Gartner}
\author[5]{J. Govaerts}
\author[4]{F.E. Gray}
\author[6]{F.J. Hartmann}
\author[2]{W.D. Herold}
\author[4]{D.W. Hertzog}
\author[1]{V.I. Jatsoura}
\author[1]{A.G. Krivshich}
\author[3]{B. Lauss}
\author[1]{E.M. Maev}
\author[1]{O.E. Maev}
\author[2]{V.E. Markushin}
\author[4]{C.J.G. Onderwater}
\author[2]{C. Petitjean}
\author[1]{G.E. Petrov}
\author[4]{C.C. Polly}
\author[5]{R. Prieels}
\author[1]{S.M. Sadetsky}
\author[1]{G.N. Schapkin}
\author[2]{R. Schmidt}
\author[1]{G.G. Semenchuk}
\author[1]{M. Soroka}
\author[1]{A.A. Vorobyov}
\author[1]{N.I. Voropaev}
\address[1]{Petersburg Nuclear Physics Institute (PNPI), Gatchina 188350, Russia}
\address[2]{Paul Scherrer Institute, PSI, CH-5232 Villigen, Switzerland}  
\address[3]{University of California Berkeley, UCB and LBNL, Berkeley, CA 94720, USA}  
\address[4]{University of Illinois at Urbana-Champaign, Urbana, IL 61801, USA}
\address[5]{Universit\'{e} Catholique de Louvain, B-1348 Louvain-La-Neuve, Belgium}
\address[6]{Technische Universit\"{a}t M\"{u}nchen, D-85747 Garching, Germany}
\address[7]{Boston University, Boston, MA 02215, USA}
  
\runningauthor{P. Kammel et al.}
\runningtitle{Muon Capture on the Proton}
% HISTORY:    %% HISTORY is optional
%\received{19 February 1997}
%\revised{}
%\accepted{}
%\dedicated{}
%\presented{}
%
\begin{abstract}
The aim of the $\mu$Cap experiment is a 1\% measurement of the singlet
capture rate \( \Lambda _{S} \) for the basic electro-weak reaction 
\( \mu +p\rightarrow n+\nu _{\mu } \). 
This observable is sensitive to the weak form-factors of
the nucleon, in particular to the induced pseudoscalar coupling constant g\( _{P} \).
It will provide a rigorous test of theoretical predictions based on the Standard
Model and effective theories of QCD.
The present method is based on high precision lifetime measurements of \( \mu ^{-} \)
in hydrogen gas and the comparison with the free \( \mu ^{+} \) lifetime. 
The \( \mu ^{-} \) experiment will be performed
in ultra-clean, deuterium-depleted H\( _{2} \) gas at 10 bar.  Low density compared to liquid 
H$_2$ is chosen to avoid uncertainties
due to pp\( \mu  \) formation. A time projection chamber acts as a 
pure hydrogen active target. It defines the muon stop position 
in 3-D and  detects rare background reactions. 
% (capture on impurities, diffusion).
Decay electrons are tracked in cylindrical wire-chambers and a 
scintillator array covering 75\% of 4$\pi$.  
\end{abstract}
\end{frontmatter}
%%%%%%%%%%%%%%%%%%%%%%%%%%%%%%%%%%%
%\def's
%%%%%%%%%%%%%%%%%%%%%%%%%%%%%%%%%%%%%
%%%*************** Text entry area *************
\section{Introduction}

The goal of this experiment~\cite{bal96,kam00,vor99} is a determination of the 
rate $\Lambda_S$ of the basic charged--current reaction
\begin{eqnarray}
\mu^- + p \to \nu_{\mu} + n
\label{mupnun}
\end{eqnarray}
via the lifetime of $\mu^-$ bound in the singlet $p\mu^-$(F=0) system
to an accuracy $\delta\tau_{\mu}/\tau_{\mu}$ better than $10^{-5}$. 
The comparison of $\tau_{\mu} = 1/\lambda_\mu$ for $p\mu$ 
atoms with that for free $\mu^+$'s will considerably improve the
determination of the 
induced pseudoscalar coupling constant 
$g_p(q^2_0=-0.88m^2_{\mu})$ to the level $\delta g_P/g_P \leq 7\%$.
The $\mu^+$ lifetime will be measured simultaneously as a reference and
to check systematics.
The lifetime measurements will be done with a time
projection chamber (TPC) filled with 10 bar of ultra-pure
deuterium-depleted hydrogen surrounded by multi wire proportional chambers
and a hodoscope of scintillation detectors. 

\begin{figure}[hbt]
{\centering \resizebox*{0.65\textwidth}{0.45\textheight}{\includegraphics{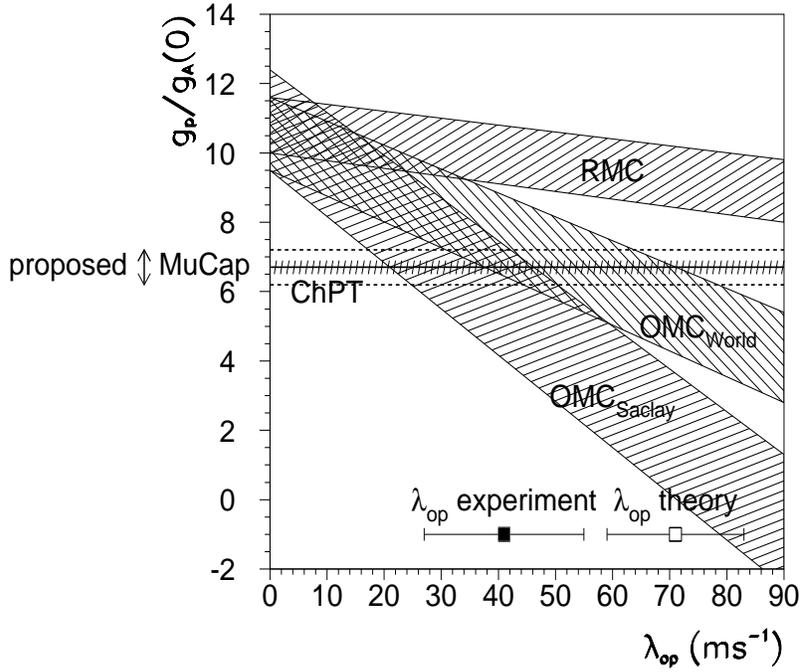}} \par}
%\mbox{\epsfysize=80mm\epsffile{muplop.eps}}
\caption{\label{FiggP}%
Current constraints on $g_P$ as function of the ortho-para 
transition rate $\lambda_{OP}$. Experimental results from 
ordinary muon capture (OMC) \cite{OMC}, 
radiative muon capture (RMC) \cite{wri98}, 
and chiral perturbation theory (ChPT).}
\end{figure}

Precision measurements of muon capture by the proton provide a challenging 
opportunity to test our understanding of chiral symmetry breaking in QCD.  
In the absence of second class currents, the electroweak structure of 
the nucleon can be described by four form factors $g_V$, $g_M$, $g_A$, and $g_P$ 
that determine the matrix elements of the charged vector and axial currents.  
While the first three of these form factors are well determined by Standard Model symmetries and experimental data,  
the pseudoscalar form factor $g_P$ is experimentally known to much less
precision (Fig.~\ref{FiggP}).   
The recent RMC result $g_P = 12.2 \pm 0.9 \pm 0.4$ \cite{wri98}
exceeds the theoretical predictions by 4.2 $\sigma$. The precision of the
older OMC results was mainly limited by absolute calibration of the 
neutron detectors. The most accurate measurement with 4.5\% 
precision was performed in Saclay~\cite{OMC} using the lifetime technique in a 
liquid hydrogen target. At this high density  $p\mu $ capture proceeds not only
from the free proton, but also from the ortho and para states of
the $pp\mu $ molecule. The uncertainty in the 
transition rate $\lambda_{op}$ between these states leads to a significant
error in the interpretation of this measurement.
A recent  experiment on $\mu^3\!\mbox{\rm He}$ capture \cite{mu3He} 
gives $g_p=8.53\pm1.54$ (the accuracy is limited by the theoretical 
extraction of $g_P$ from the three--nucleon system) in better 
agreement with theory. 

This controversial experimental situation is in stark contrast to the recent
progress in the theoretical understanding of muon capture achieved within 
the framework of low energy effective theories of QCD.
The dominant contribution to the pseudoscalar form factor 
is given by the pion pole (PCAC), and the leading correction to the pole 
term can be derived from QCD Ward identities \cite{ber94} confirming the 
old current--algebra result \cite{AD}.
Possible higher order corrections appear to be small \cite{ber98}.
Recent calculations of singlet and triplet muon capture rates
in heavy--baryon chiral perturbation theory\cite{AMK2000}  
and in the small-scale expansion \cite{BHM00} are in a good agreement with 
the careful analysis \cite{gov00}. 
\vspace{.2cm}

\begin{center}
\begin{tabular}{clllll}
\hline
  Reference              &\cite{ber94} &\cite{fea97} & \cite{gov00}   & \cite{AMK2000}  & \cite{BHM00} 
\\
\hline 
  $g_P$  & $8.44\pm 0.23$ & $8.21 \pm 0.09$  & $8.475 \pm 0.076$   &   & \\        
\hline 
  $\Lambda_S\;\;(\mathrm{s}^{-1})$ & &         & $688.4 \pm 3.8$     & $695$   &   $687.4$\\         
\hline 
  $\Lambda_T\;\;(\mathrm{s}^{-1})$ & &         & $12.01 \pm 0.12$     & $11.9$  &  $12.9$ \\        
\hline
\end{tabular} 
\end{center}
\vspace{.2cm}

The main experimental challenges for a significantly improved measurement of $\Lambda_S$ 
result from three different sources.
a) statistics: The capture rate will be determined as 
\( \Lambda _{S}= \)\( \lambda _{\mu -}-\lambda _{\mu +} \),
thus \( \delta \Lambda _{S}\sim  \)\( \sqrt{2}\ \delta \lambda _{\mu } \). As $\Lambda _{S} 
\approx 1.5 \times  10^{-3} \lambda _{\mu} $ the lifetime of the positive and negative muon have 
to be measured with at least 10 ppm precision, i.e. 10$^{10}$ reconstructed events each.
b) interpretation:
$pp\mu $ formation should be slow so that capture 
takes place predominantly from the well defined F=0 hfs states of the muonic hydrogen
atom. 
c) distortions of the muon decay time spectrum:
Dangerous physics effects for $\mu^-$ include direct or delayed muon stops in the 
wall material, transfer to gas impurities or to deuterium. 
The $\mu^+$ polarization is
partially preserved in hydrogen, leading to effects of muon spin rotation and
relaxation. Detector imperfections might cause time dependent efficiency variations
due to instrumental correlations between a muon and its decay electron and overlapping
muon--electron pairs.  

\section{Experimental set-up}

The central part of our detector is a time projection chamber (TPC)
embedded in a pressure vessel filled with 10 bar of ultra-pure
deuterium-depleted hydrogen (protium). The TPC which was specially
developed for this experiment has a sensitive volume of 15 x 12 x 30 cm$^3$
and acts as active target monitoring all muon stops and electrons
from muon decay. 
The vertical drift field 
of $\sim$2.4~kV/cm causes electrons 
to drift with a velocity of $\sim$ 0.7 cm/$\mu$s toward multiwire proportional
planes at the bottom. There, charges are amplified by typically a
factor of 10$^4$ and read out by 75 anode wires in x-direction and by 38 
cathode strips made of wires in z-direction.
The y-coordinate, defining the height in the TPC, is determined
by the drift time which ranges from 0 to 17 $\mu s$.
Incoming muons are detected by two planes of wire
chambers in front of the TPC.
Track reconstruction inside the TPC clearly distinguishes between muons stopping in H$_2$ versus 
in the walls,
thus  allowing the use of low gas density (1\% of liquid H$_2$) which reduces
problem b) to a negligible level. 

The pressure chamber has cylindrical walls made of 4 mm aluminum
to reduce multiple scattering of
through-going decay electrons. 
The hydrogen vessel and its interior wire chambers
are made of clean materials (metals, ceramics, quartz-glass frames, etc.) that
can be baked out up to 150$^o$C and evacuated down to $10^{-7} - 10^{-8}$ mbar.
This level is required to maintain a required hydrogen purity of 10$^{-8}$.
Ultra clean protium is filled via a specially developed gas system
using chemical purifying methods. The gas can be circulated and purified
during the measurements. Since this is an active target experiment, very
low levels of impurities can be determined from the chamber signals 
themselves, in addition to chromatographic gas analysis.

Surrounding the pressure tank, two cylindrical proportional chambers 
and an array of plastic detectors are mounted covering 
an effective solid angle $\Omega /4\pi\sim$75\%.
For the electron time determination, the measurements will
rely entirely on the detectors outside the hydrogen pressure vessel, i.e.
on the two wire chambers for directional back tracking and on the plastic
hodoscope for the absolute time measurement. The separation of detector
functions for electrons from those for muons ensures independent absolute
time measurements without the danger of electronic cross-talks and tail effects.
The tracking chambers can handle event rates
of $\sim$ 30 kHz, since pile-up problems can be 
reduced by identifying the muon-electron pair originating from a 
common vertex. This method also suppresses other possible background.  

\begin{figure}[hbt]
\begin{center} 
{\centering \resizebox*{.8\textwidth}{0.4\textheight}
{\includegraphics{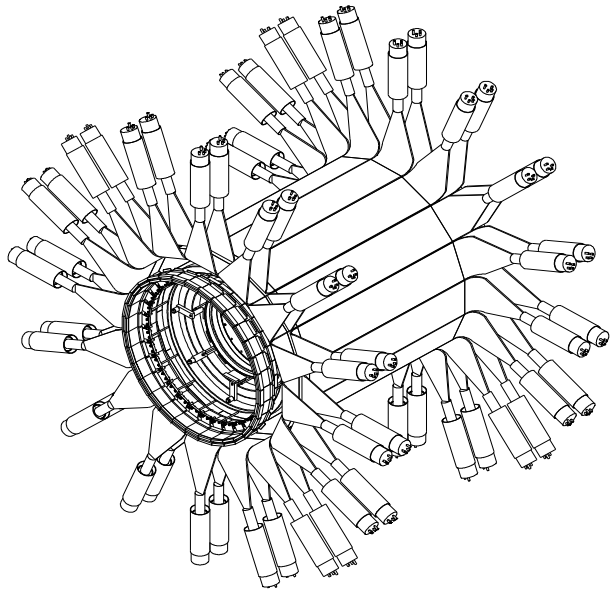}} \par}
%{\includegraphics{Hinten_Rechts.eps}} \par}
%\mbox{\epsfysize=80mm\epsffile{TPC.eps}}
\caption{\label{TPC}
Overview of $\mu$Cap detector showing scintillator array and endframes of 
the cylindrical wire chambers used for electron tracking.
The hydrogen vessel with muon chambers and TPC is located inside the electron detector
system.}
\end{center}
\end{figure}
\vspace{-5mm}
A coil will be installed outside the TPC vessel to generate a uniform magnetic
field of $\sim$70 Gauss tranverse to the beam and muon spin axis. This field will
precess the remaining free muon polarization at 1 MHz creating a
sinusoid on top of the exponential decay spectrum. Monte Carlo studies indicate
that amount and relaxation of this polarization can be determined in
a fit over 15 $\mu$s. The oscillating part largely
decouples from the lifetime measurement.
Moreover spin asymmetries get strongly reduced by the
cylindrical symmetry of the setup.

%Inner part of the final experimental setup (flanges of pressure tank,
%entrance chambers and TPC) in stereoview from the back side}

\section{Detector performance}

\begin{figure}[hb]
{\centering \resizebox*{.9\textwidth}{0.51\textheight}
{\includegraphics{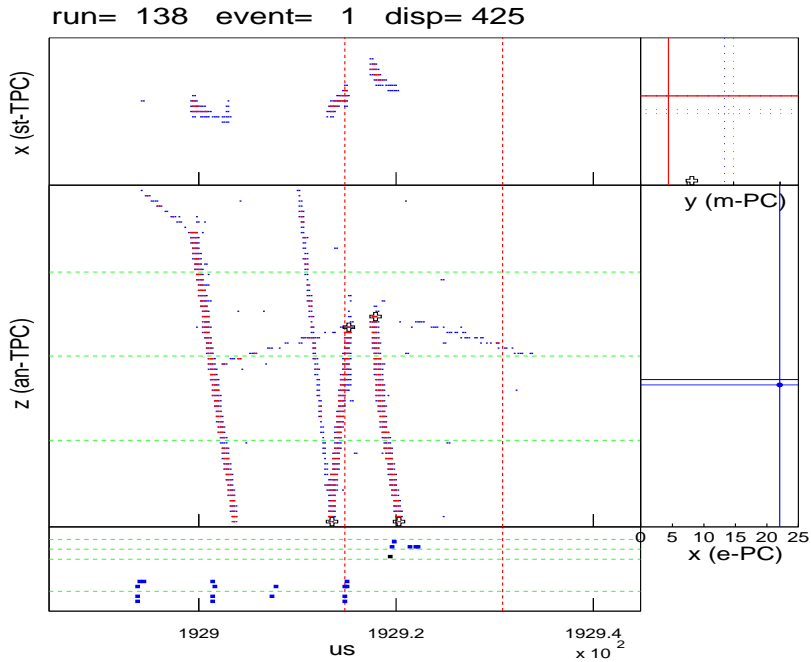}} \par}
\vspace{-.0cm}
\caption{Event display of 60~$\mu$s time slice.  The central panel (y-z plane) shows
TPC anode signals as function of time (\protect\( \mu \protect \)s). Muons are distinguished
from electrons by using two discriminator levels. The upper panel (y-x plane)
indicates the corresponding information from the strip cathodes.
The bottom panel shows the absolute times of various detector hits. 
The right panels indicate wires hits on muon (upper) and electron (lower)
MWPC's, respectively. 
\label{save138}}
\end{figure}

Significant
R\&D was required to establish the feasibility of the new method and to optimize physics
and detector parameters. In particular, several engineering runs with a prototype
TPC have been performed at PSI (cf. contribution~\cite{mae01} and 
references~\cite{bal96,kam00,vor99}). 
The performance of the prototype detector was satisfactory and proved that stable chamber
operation in pure hydrogen can be achieved. Rather high
chamber voltages  of 6.8~kV on 2-4~mm spacing were required in order to
obtain sufficient gas amplification.
The main data was recorded by a custom-designed dead-time free TDC (TDC400) operating
at a clock rate of 5~MHz. The hits of all detectors were stored for
contiguous time regions of $\sim$10~ms, providing the full history
information around individual muon stops. A short time slice from this
time region is displayed in Fig.~\ref{save138}. 
\vspace{-.5cm}

\begin{figure}
{\centering \resizebox*{0.6\textwidth}{0.4\textheight}
{\includegraphics{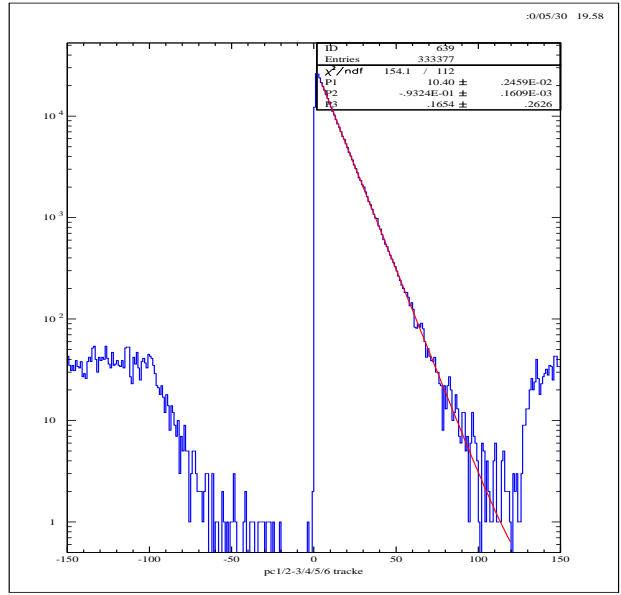}} \par}
\caption{Time distribution of fully tracked $\mu$-e events with global pile-up
protection. The accidental level is below 10$^{-5}$ of the signal at time 0.
\label{e1_pu_xyz}}
\end{figure}

The behavior of the observed $\mu^-$ decay rate
is consistent with the decline of the target purity with time and purity recovery
by refilling. The \( \mu ^{+} \) spectra reveal \(  \)\( \mu ^{+} \)SR effects
consistent with muonium precession in
a residual magnetic field of 0.2-0.4 G (no tranverse field was applied during this test).
For pile-up free events the accidental to signal level is already
$\sim 10^{-4}$ for a simple coincidence between a MWPC and electron telescope.
It can be further improved by an order of magnitude using the 4 MWPC's and the TPC 
to precisely track the decay electron back to the end of the muon track (Fig.~\ref{e1_pu_xyz}). 

By using the TPC as active target, charged products of
muon induced reactions with impurities can be detected. Most critical are 
O\( _{2} \), N\( _{2} \), H\( _{2} \)O and D\( _{2} \). The TPC is sensitive to
recoil nuclei (200-350 keV) from \( \mu  \)-capture on impurities with Z$>$1
and from the $pd \mu $ fusion channel \( ^{3}He (0.2 MeV)+\mu (5.3 MeV) \).
The information is collected both with the TDC system and from 12 TPC anodes
instrumented with FADC's. 
For the selection of  \( \mu  \)-capture reactions logarithmic amplifiers, 
discriminators with high threshold about 70~keV and a hardware trigger were 
developed~\cite{mae01}.
For about 10$^6$ muon stops in the sensitive  TPC
volume 3876 $\mu$-capture
events with $\mu^-$ beam and only one event under the same requirements
with $\mu^+$ beam were found. As this data corresponds to an
impurity level of 30~ppm as estimated by chemical
analysis, a sensitivity to determine impurities with Z$>$1 of about 0.01~ppm has 
been demonstrated. 

\begin{figure}[hbt]
\begin{center} 
{\centering \resizebox*{0.9\textwidth}{0.4\textheight}
{\includegraphics{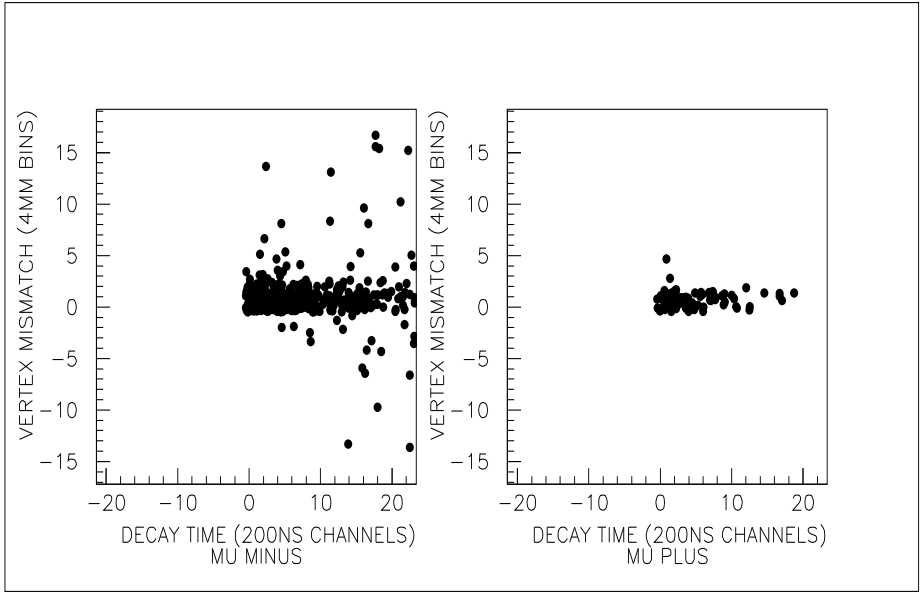}} \par}
%put the diffusion study plots in here.
%\mbox{\epsfysize=80mm\epsffile{TPC.eps}}
\caption{\label{MUDIFF2D}%
Distance between muon stop and decay point as function of decay time
from $\mu-e$ tracks reconstructed inside the TPC.
$\mu^-$ data~(left), $\mu^+$ data~(right). Significant diffusion was observed
for $\mu^-$ after d$\mu$ formation.
}
\end{center}
\end{figure}

Due to the  \char`\"{}Ramsauer-Townsend\char`\"{}
minimum in elastic d$\mu$-p scattering, d$\mu$ atoms formed after transfer from protium
can travel several cm in 10 bar hydrogen. Some fractions even leave the sensitive volume
of the chamber, especially at late times after muon stop. Therefore, the \( \mu  \)-e
time distribution can get significantly distorted toward a steeper slope simulating
a larger capture rate. This is the principal reason why it is necessary to use
protium which is strongly depleted from any deuterium.
The distortions  depend on the geometrical $\mu-e$ vertex cuts used, increasing with
tighter cuts. 
Fortunately the diffusion correction can be self calibrated 
by monitoring deuterium
impurities in the experiment  and then extrapolating to zero deuterium concentration.
The most promising method is the direct observation of muon diffusion by observing an electron track
displaced from the muon stop point in the TPC.
Fig.~\ref{MUDIFF2D} shows the result of a d$\mu$ diffusion search in 
our test runs.   The outlying points in the left image are from the
diffused d$\mu$ atoms in a $\mu^-$ run.  For reference only
a single event at t=0 leaks through in a $\mu^+$ run shown
on the right.  
Note the importance of the intrinsic TPC tracking accuracy and 
electron efficiency for this analysis.

\section{Summary and outlook}

We plan to analyze the data with two main and complementary methods.\\
{\em global pile-up free data}. Only events are analyzed which do not have another muon
entering the TPC within 10-20 $\mu$s.
This analysis is particularly clean and simple, and thus 
ideally suited for a precision experiment. Moreover it allows loose vertex cuts, thereby reducing
the experimental sensitivity to deuterium diffusion. It has the disadvantage
of significant pile-up losses.\\
{\em local pile-up free data}. The TPC volume is subdivided
in smaller volumes during the analysis by requiring no other local muon stop
inside a cut volume around the electron reconstructed vector. In this way events
can be analyzed where several simultaneous muons stop during the measurement/drift-time
interval of the TPC. The vertex matching suppresses the accidental background
from uncorrelated muons, but systematic corrections become more important.\\ 
The choice between the two analysis methods is made off-line, as in our experiment
contiguous event-blocks containing all information are recorded. 
A summary of estimated statistical and systematic corrections and errors in ppm of 
the measured lifetimes is given below.
\vspace{0.2cm}

\begin{small}
\centering \begin{tabular}{ccccccc}
\hline 
\textbf{}&
\multicolumn{3}{c}{\textbf{global PU free data}}&
\multicolumn{3}{c}{\textbf{local PU free data}}\\
\hline 
\textbf{}&
\textbf{\( \mu ^{-} \)}&
\textbf{\( \mu ^{+} \)}&
\textbf{comment}&
\textbf{\( \mu ^{-} \)}&
\textbf{\( \mu ^{+} \)}&
\textbf{comment}\\
\hline 
{\bf statistics}&
{\bf(10)}&
{\bf(10)}&
10\( ^{10} \)\ events&
{\bf(7)}&
{\bf(7)}&
2x10\( ^{10} \)\ events \\
\hline 
wall stops&
(2)&
-&
&
(2)&
-&
\\
\hline 
impurities&
2(3)&
-&
for c\( _{Z} \)=10\( ^{-8} \)&
2(3)&
-&
for c\( _{Z} \)=10\( ^{-8} \)\\
\hline 
flat acc.&
(2)&
(2)&
level 10\( ^{-4} \)&
(3)&
(3)&
level 5x10\( ^{-4} \)\\
\hline 
\( \mu  \)SR&
-&
(2)&
&
-&
(2)&
\\
\hline 
diffusion&
1(1)&
-&
no vertex cut&
100(5)&
-&
5 cm radial cut\\
\hline 
two event corr.&
- &
-&
&
(2)&
(2)&
acc. structure\\
\hline 
\textbf{total sys. error}&
\textbf{4.2}&
\textbf{2.8}&
\textbf{}&
\textbf{7.1}&
\textbf{4.1}&
\textbf{}\\
\hline
\hline 
\textbf{\( \delta \lambda  \) total error}&
\textbf{10.9}&
\textbf{10.4}&
\textbf{}&
\textbf{10.}&
\textbf{8.1}&
\\
\hline 
\end{tabular}
\end{small}
\vspace{0.2cm}

The final $\mu$Cap detector is presently under construction and first
production running is expected in fall 2002. There exist several exciting
future extensions of the physics reach for this program. A second phase of the experiment
might aim at a 0.3\% measurement of $\Lambda_S$ to achieve an experimental
precision of 3\% in $g_P$ matching the present theoretical error. This proposal
is based on an intense chopped muon beam~\cite{kam98}, which is presently being developed for
the $\mu Lan$ experiment~\cite{car99}. 
A very recent idea~\cite{kam01} concerns a precision measurement
of the d$\mu$ capture process, both its integral rate and Dalitz Plot
distribution. Within the framework of pion-less effective theories muon capture
on the deuteron is directly related to the  pp fusion in the sun~\cite{che01}
and neutrino deuteron scattering as observed in the Sudbury Neutrino Observatory~\cite{sno01}
and, thus, appears a unique possibility to calibrate these fundamental processes
of current topical interest.

Finally, let me express my sincere gratitude to Ken Nagamine and the organizers of MCF'01 for
the opportunity to enjoy this stimulating meeting and let me congratulate him for his
remarkable  contributions to this field of muon physics.

\end{document}